\documentstyle[12pt,epsfig]{article}
\newfont{\prg}{cmsy10}

\begin{document}
\phantom{}
\vspace{10mm}
\begin{center}
{\Large  Saturation regimes at LHC energies
}  \\
\vspace{10mm}

{J.-R. Cudell, O.\,V. Selyugin\footnote{
BLTPh, JINR, Dubna, Russia.   } \\
{\small \it Institut de Physique,
Universit\'e de Li\`ege,
Belgium} }
\%today{empty}

\end{center}

\vspace{10mm}

\begin{abstract}
  The effects of saturation and unitarization for hadron-hadron scattering
at LHC energies are
   explored in several models.
It is shown that different choices of saturation parameters
lead to sizable differences in the energy dependence
of total cross sections, and to dramatic changes in elastic
differential cross sections.
\end{abstract}

\section{Introduction}

Saturation is now a very popular term, and it has a very
wide meaning. It includes ``shadowing'' and ``anti-shadowing'' processes,
 gluon saturation, unitarization, etc.
In this paper, by ``saturation'' we mean that the ``black disk''
limit (BDL) has been reached, and that effects connected to the
unitarity of the scattering amplitude must be taken into account.

Standard ingredients for the discussion of this regime can be found, e.g.
in the loop-loop correlation model \cite{sochi},  which is based
on a functional integral approach to high-energy
collisions \cite{nach}. Here,
the $T$-matrix element for elastic
scattering is given by
the correlation function of two light-like Wigner-Wilson loops:
\begin{eqnarray}
        T_{pp}(s,t)
        & = & \,\,
        2 i s \ \int  \ d^2b_{\!\perp}\,
        e^{i {\vec q}_{\! \perp}\cdot {\vec b}}\,
        J_{pp}(s,|\vec{b}|)
\label{Eq_T_pp_matrix_element} \nonumber \\
{\rm with\ }        J_{pp}(s,|\vec{b}|)
        & = &
        \int \!\!dz_1 d^2r_1\!\! \int \!\!dz_2 d^2r_2
        |\psi_p(z_1,\vec{r}_1)|^2 |\psi_p(z_2,\vec{r}_2)|^2
\nonumber\\ &&
        \times
        \left[1-S_{DD}(s,{\vec b},z_1,{\vec r}_1,z_2,{\vec r}_2)
      \right]
\label{Eq_model_pp_profile_function}
\end{eqnarray}
The saturation of the profile function $J_{pp}$ appears then as a
manifestation of the unitarity of the $S$-matrix.

Whether the BDL will affect LHC physics, and how it is reached in QCD
is still a matter of debate. For instance, if we consider its effect on
the total cross section
$\sigma_{tot}$
(which can be obtained directly from $T_{pp}$),
the standard point of view is that
the saturation of
$J_{pp}(s,|\vec{b}|)$ tames its growth:
there is a transition from a power-like to an
$\ln^2 (s)$-increase of $\sigma^{tot}_{pp}(s)$, which then respects the
Froissart bound

But where this effect will take place is not clear:
the analysis of \cite{sochi} predicts that the saturation regime
at small $b$ is reached only at very small $x \approx 10^{-10}$ and very high
energies $\sqrt{s} \geq 10^{6} \ $GeV.
Thus -- according to this model -- the onset of the BDL
in $pp$ collisions is about two orders of magnitude beyond
LHC energy.
However, other authors (see e.g. \cite{bart}) predict the same effect,
within the dipole picture of soft processes,
but at much lower energies.

The connection with unitarisation becomes clear if we notice that
the eikonal representation for the scattering amplitude
in $\vec b$-space, in the form $1-\exp(-\chi(s,b))$, reaches
the BDL only asymptotically. However, this representation is not the
only possibility, and it may be more useful to consider the effects
of saturation by
considering parametrisations
in $s$ and $t$, transforming them to impact parameter space, and
imposing the BDL as an upper bound on the amplitude in $s$ and $b$.
\section{   Donnachie -Landshoff (DL) model}
Hence, we shall first consider this approach in the DL model,
which must certainly be unitarized at high energies.
Here, the $pp$-elastic scattering amplitude is
proportional to the hadronic form factors, and can be approximated
at small $t$ by:
\begin{eqnarray}
 T(s,t) \ =  \ h_{0} \ s^{\epsilon}
 e^{D \ t} e^{\alpha{\prime} \  t \ \ln s}.
\end{eqnarray}
Going to impact parameter space, we obtain the 1-pomeron amplitude
$\Gamma_1(b,s)$:
\begin{eqnarray} 
 \Gamma_1(b,s)  = \ h_{0} \ \frac{(s/s_0 )^{\epsilon}}{2 \
 [4 + \alpha^{\prime} \ln{s/s_0}]}
 \ \exp\left(-{b^2\over4(4 + \alpha^{\prime} \ln{s/s_0})}\right),
\end{eqnarray}
with $h_0=4.7 \ $GeV$^{-2}$; $\epsilon=0.0808$;
$\alpha^{\prime}=0.25$ GeV$^{-2}$; $D=4$ GeV$^{-2}$ and $s_0 =1$ GeV$^2$ ($s_0$ will be
dropped below).
Double-pomeron exchange then gives:
$$ \Gamma_2(b,s) \ = \ \Gamma_1 (b,s) - \lambda \Gamma_1^2 (b,s)/2 $$
At some energy and at small $b$, $\Gamma_1(b,s)$ and $\Gamma_2(b,s)$ reach
 the  BDL
(see Fig.~1). For one-pomeron exchange, this happens at $\sqrt{s}=1.5 \ $TeV
whereas the 2-pomeron exchange term moves the limit up to
$\sqrt{s}=4.5 \ $TeV (for $\lambda=0.2$).

The details of the saturation process are very important
at super-high energies. In Fig.~2, we show different ways to implement
saturation at $\sqrt{s}=14 \ $TeV:
the solid line shows $\Gamma_1$ without saturation;
the dashed line corresponds to the ``hard cut'' case,
where we simply freeze $\Gamma_1$ at 1,
which corresponds to a specific point
where the curve has a knee; the dash-dotted curve and the circles represent
the above cases for $\Gamma_2$. Simply freezing $\Gamma$
breaks analyticity, hence we must consider scenarios where the BDL is smooth.
First of all, we can continue the $\Gamma=1$ plateau by a
Gaussian which connects quickly to the original curve (``soft cut'').
Secondly, we shall consider a (``scaling'') case: once $\Gamma$ reaches 1, we
assume that the factor multiplying the exponential in Eq.~(3) is
renormalised to 1, but the curve gets shifted towards higher
values of $b$, such that
the growth of the radius (defined as the point at which $\Gamma=1/2$)
remains the same as in the unsaturated curve.
This gives us a smooth growth of the saturated core, as shown by
the dotted curve in Fig.~2. Although it slightly changes
the form of $\Gamma$ near the knee, it gives the same asymptotic
behaviour at large $b$.

Fig.~3 shows the consequences of the above saturation schemes on $\sigma_{tot}$.
The solid line corresponds to the case without saturation,
the dash-dotted line to the ``hard cut'', the
dashed line to the ``soft cut''  and the dash-dotted-dotted line to
the ``scaling'' case.
  Fig.~3 also gives the results
of full eikonalization ($\Gamma(b,s) = 1-\exp[-1.233\ \Gamma_1(b,s)]$)
as a long dashed line and of the Dynamic Peripheral Model (DPM) \cite{gks} as
a dotted line.  We see that $\sigma_{tot}$ does not depend much on
the details of saturation at LHC energies, whereas
at very large energies a sizable fraction of the cross section will
come from these, and hence from the hadron structure at
large distances.


\begin{figure}{t!}
\vspace{-4.cm}
\begin{flushleft}
\phantom{.}\hspace{-5mm}
\mbox{\epsfysize=60mm\epsffile{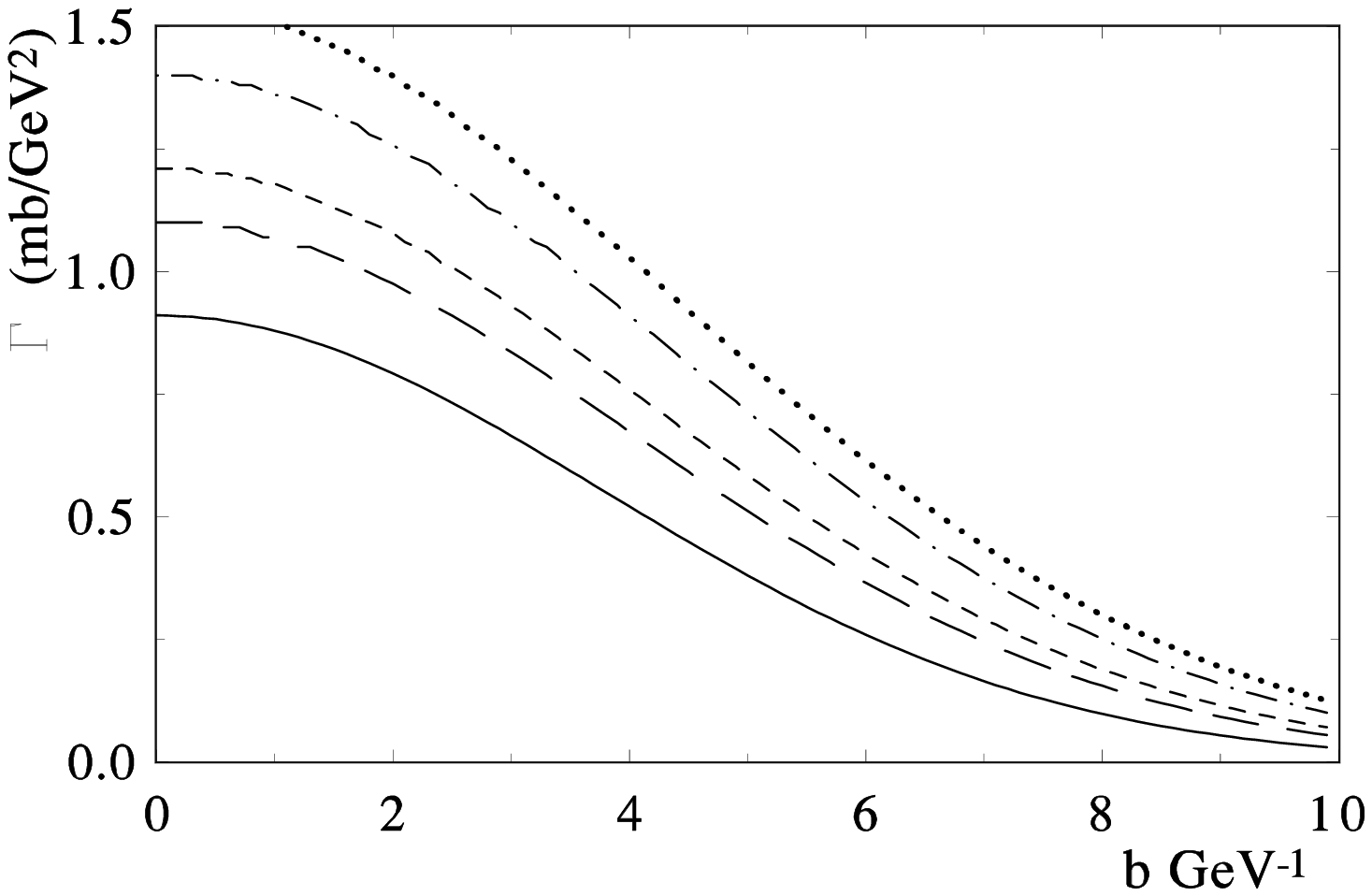}}
\end{flushleft}
\vspace{-0.5cm}
{\small Fig.~1: The $s$ and $b$ dependence of $\Gamma_1$
 }

\vspace{-7.8cm}

\begin{flushright}
\phantom{.}\hspace{60mm}
\mbox{\epsfysize=60mm\epsffile{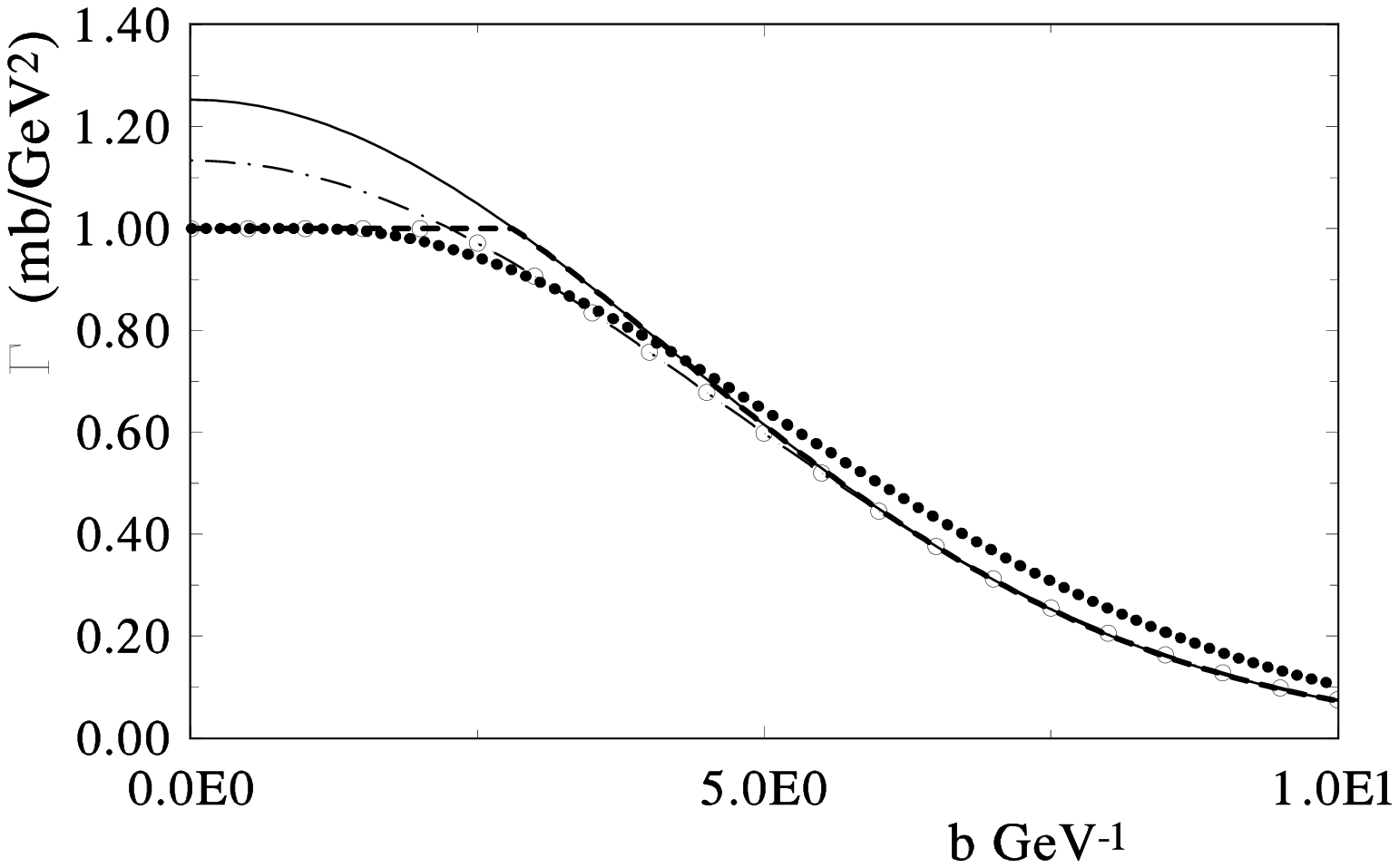}}
\begin{minipage}{5.7cm}
{\small Fig.~2:  $\Gamma_1$ and $\Gamma_2$ at $14 \ $TeV \\
   (see explanation in text)}
\end{minipage}
\end{flushright}
\end{figure}

\vspace{0.4cm}

\vspace{-7mm}

\begin{figure}
\vspace{-2.cm}
\begin{flushleft}
\phantom{.}\hspace{-5mm}
\mbox{\epsfysize=60mm\epsffile{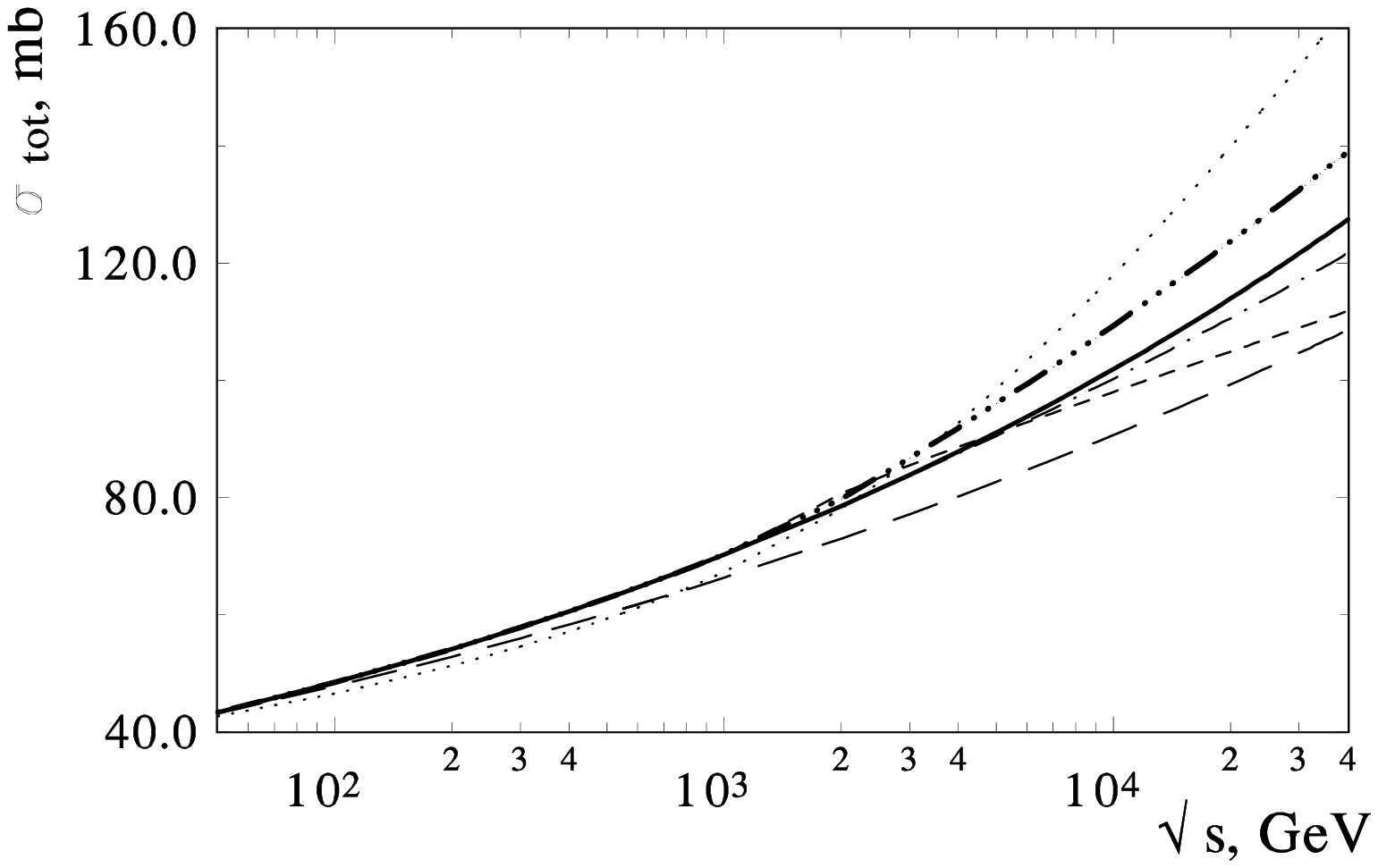}}
\end{flushleft}

\vspace{-7.8cm}

\begin{flushright}
\phantom{.}\hspace{60mm}
\mbox{\epsfysize=60mm\epsffile{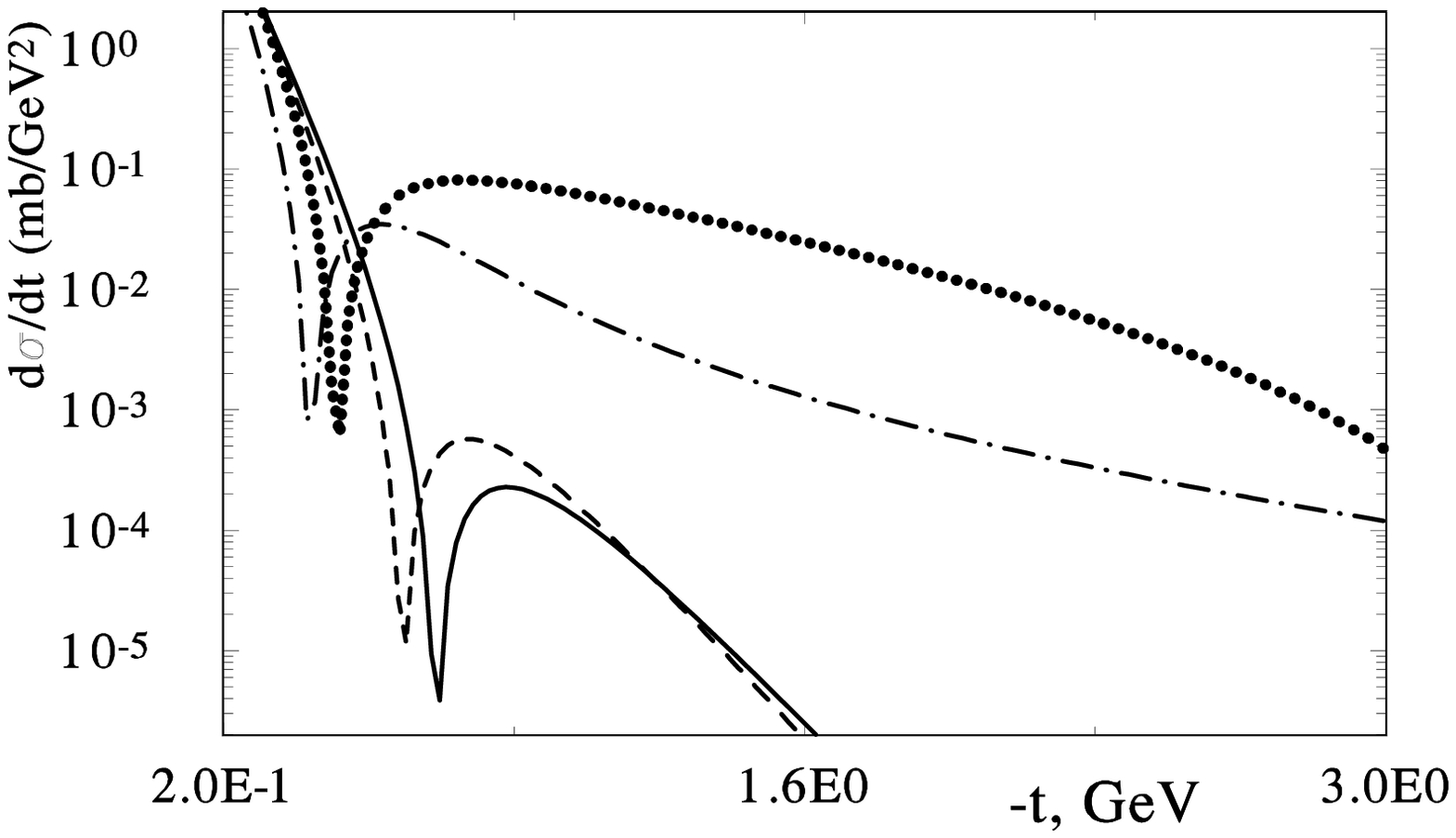}}
\end{flushright}
\end{figure}

\vspace{2.5cm}

\begin{minipage}{5.7cm}
{\small Fig.~3: The different predictions of  \\ $\sigma_{tot}$
 (see explanation in text)}
\end{minipage}

\vspace{-15mm}
\begin{flushright}
\begin{minipage}{5.7cm}
{\small Fig.~4:  $d\sigma/dt_{pp}$ in the $pp$ case, at 4 and 14 TeV.}
\end{minipage}
\end{flushright}

\vspace{1.cm}

At the LHC, the effects of saturation can however be observed
in elastic scattering.  Fig.~4 shows that the saturation
effects can lead to a large increase in the differential cross section
at large $|t|$.
The solid and long-dashed lines represent $d\sigma/dt$ at $4$ and $14 \ $TeV,
without
saturation. The dash-dotted line and the points show $d\sigma/dt$ at
$14 \ $TeV in the ``hard cut'' and ``scaling'' cases. Both scenarios,
while giving different predictions for the growth of $\sigma_{tot}$,
lead to a similar large growth of the differential cross section. Note
that  we do not aim here at quantitative predictions, but only want
to gauge possibilities.

\section{ The Dynamic Peripheral Model}
We can compare the above results with those of the eikonal
dynamical model of $hh$-scattering developed in \cite{gks}.
The model is based on the general principles of quantum field theory
\vspace{0.5cm}

\begin{figure}
\vspace{-3.3cm}
\begin{flushleft}
\phantom{.}\hspace{-5mm}
\mbox{\epsfysize=60mm\epsffile{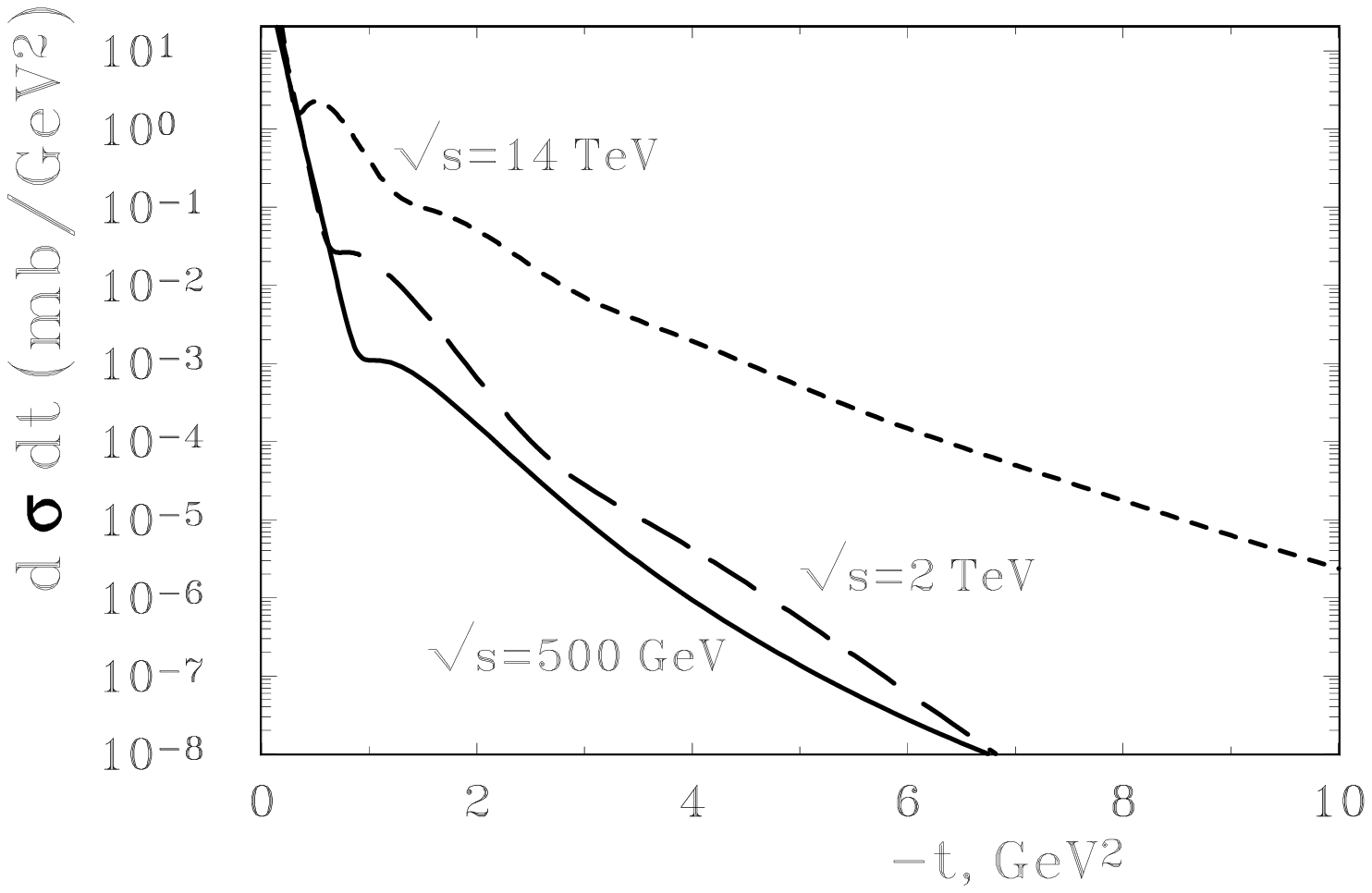}}
\end{flushleft}

\vspace{-6.7cm}

\begin{flushright}
\mbox{\epsfysize=60mm\epsffile{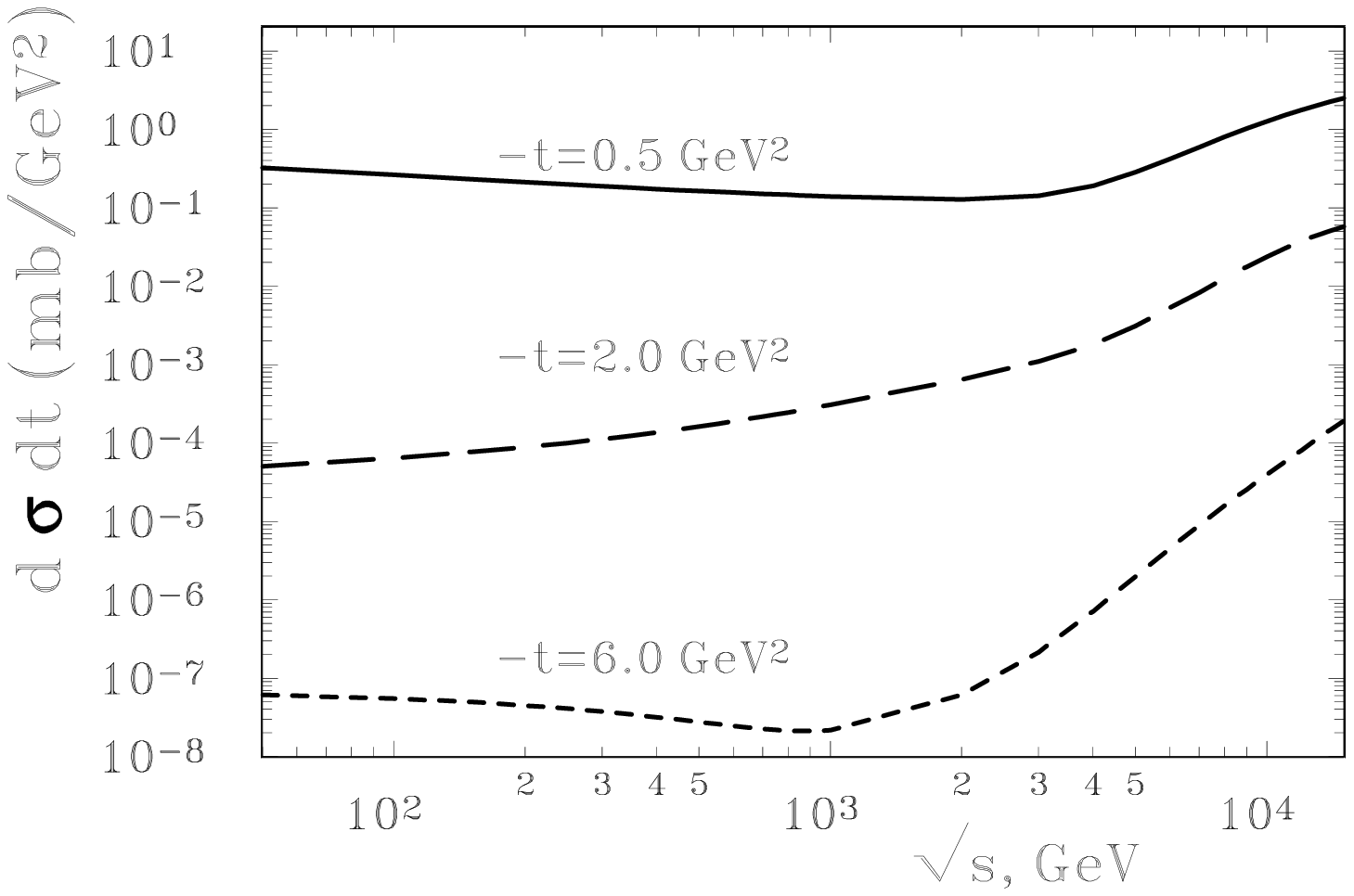}}
\end{flushright}
\end{figure}


\begin{flushleft}
\begin{minipage}{7.0cm}
\vspace{-1.cm}
\phantom{.} \hspace{-0.2cm} {\small Fig. 5:
$d\sigma/dt$ from the DPM in the $pp$ case.}
\end{minipage}
\end{flushleft}

\begin{flushright}
\vspace{-1.5cm}
\begin{minipage}{5.5cm}
{\small Fig.~6:
$d\sigma/dt$ from the DPM in the $pp$ case
  at fixed $t$.}
\end{minipage}
\end{flushright}

(analyticity, unitarity, causality ) and takes into account basic
information on the structure of a hadron as a compound system, with
a central region in which the valence quarks are concentrated and
with a long-distance region occupied by the color-singlet quark-gluon field.
The model provides  a self-consistent  picture  of
$d\sigma/dt$ and of spin phenomena for different hadron processes.

  The predictions of the DPM for elastic $pp$-scattering at
$\sqrt{s} = 53 \ $GeV  and $14 \ $TeV are shown in Fig.~5.
At high energies, on can see that $d\sigma/dt$
changes its behavior and grows with increasing
energy.
This can be seen in more details in Fig.~6.
This effect is due to the peripheral term
 which comes from the interaction of the pomeron with the two-pion cut
and leads to a growth of $\sigma_{tot}$ as
 $\sim (\ln{s})^2$ and to a rapid growth of $d\sigma/dt$ at large $|t|$.
It is evident that the
 growth of the size of proton increases the role of
  peripheral effects at super-high energies,  where
 the BDL is reached.
 After that,  $d\sigma/dt$ changes behavior and begins
 to grow at fixed $t$.
\section{Conclusion}
We have shown that the saturation effects, in very different
  approaches, may be observable at the LHC. They
   lead on the one hand to a change in the monotonic behavior
  of $\sigma_{tot}$
  and, on the other, to an increase of   $d\sigma/dt$ at large $|t|$.
Hence the LHC experiments may tell us if we reached
the saturation regime, and provide information on
the hadron dynamics at large distances.

\bigskip

{\small The authors would like to thank P. Landshoff for helpful discussions.
    O.V.S. is a Visiting Fellow of the
Fonds National pour la
Recherche Scientifique, Belgium.}

\end{document}